\documentclass[showpacs,aps,prl,twocolumn,superscriptaddress]{revtex4-1}

\usepackage[english]{babel}
\usepackage{graphicx}
\usepackage[colorlinks=true,allcolors=blue]{hyperref}
\usepackage{float}
\usepackage{xcolor}

\begin{document}

\def\equationautorefname~#1\null{Eq.~(#1)\null}
\renewcommand{\figureautorefname}{Fig.}

\title{Chiral Discrimination in Helicity-Preserving Fabry-P\'erot Cavities}

\author{L. Mauro}
\thanks{These two authors contributed equally.}
\affiliation{Univ. Bordeaux, CNRS, LOMA, UMR 5798, F-33400 Talence, France}
\author{J. Fregoni}
\thanks{These two authors contributed equally.}
\affiliation{Departamento de Física Teórica de la Materia Condensada and Condensed Matter Physics Center (IFIMAC), Universidad Autónoma de Madrid, 28049 Madrid, Spain}
\author{J. Feist}
\email{johannes.feist@uam.es}
\affiliation{Departamento de Física Teórica de la Materia Condensada and Condensed Matter Physics Center (IFIMAC), Universidad Autónoma de Madrid, 28049 Madrid, Spain}
\author{R. Avriller}
\email{remi.avriller@u-bordeaux.fr}
\affiliation{Univ. Bordeaux, CNRS, LOMA, UMR 5798, F-33400 Talence, France}

\date{\today}

\begin{abstract}
We theoretically study circular dichroism of chiral molecules embedded inside an helicity-preserving Fabry-P\'erot cavity.
We find an increase of the intrinsic chiroptical response of the molecules by two orders of magnitude, and report the first clear signature of \textit{chiral cavity polaritons} upon entering the regime of strong light-matter coupling.
We study a cavity design based on two dielectric photonic crystal mirrors acting, in a narrow frequency range, as efficient polarization cross-converters in transmission for one polarization, and almost perfect reflectors for the other polarization.
%
%
We show that a Pasteur medium hosted inside such a cavity can couple efficiently to both the outside of the cavity and to the helicity-preserving mode, inheriting an enhanced chiral character.
We expect such a device to be useful in the future to design ultrasensitive chiral sensors for optics and stereochemistry. 
\end{abstract}
\maketitle
%

%
%
Optical activity (OA) and circular dichroism (CD) are fundamental optical effects depending on the polarization of light~\cite{landau2013electrodynamics,Condon1937,bruhat1992optique}.
%
%
%
Analysis of OA and CD signals has a wide range of applications in physics~\cite{Hecht1994,Craig1998,Barron2004,Tang2010,Lozano2018}, stereochemistry~\cite{snatzke_circular_1968} and biology~\cite{beychok1966circular,ranjbar_circular_2009}, in particular as a spectroscopic tool to discriminate between chiral enantiomers, molecules that are non-superimposable mirror-images of each other.
%
In the simplest case of a homogeneous, linear, and isotropic (HLI) medium, OA is due to an off-diagonal coupling between the medium electric (magnetic) polarization and the magnetic (electric) component of the propagating field~\cite{Condon1937,landau2013electrodynamics}.
%
%
%
%
This coupling is given by the dimensionless Pasteur pseudo-scalar coefficient $\kappa$, which is usually very small (of order $\approx 10^{-3}-10^{-5}$~\cite{Condon1937,Craig1998}), thus implying intrinsically weak OA and CD signals, and limiting the sensitivity of traditional spectroscopy to discriminate low concentrations of chiral molecules.
%
%

%
%
Several approaches have been proposed to circumvent this problem.
A fruitful one has been to design and fabricate 2D metasurfaces~\cite{Oh2015,Papakostas2003,Zhang2009,Decker2009,Li2013,solomon_enantiospecific_2019}, and planar chiral plasmonic structures~\cite{Kuwata2005,Collins2017}, in order to engineer transmission and reflection properties of their scattered light~\cite{menzel_advanced_2010}.
This improved the ability of such materials to enhance the chiral response of molecules deposited on their surface~\cite{Choi2013,Corbaton2019,Rho2020}, and thus to perform improved chiral sensing~\cite{Park2019,Chen2022}.
%
Another successful route has been to deposit molecules on the surface of nanoparticles and induce a plasmon-enhanced CD~\cite{Govorov2013,Weiss2016,Wang2019,Vestler2019,Lieberman2008}.
%
%
%
\begin{figure*}[tb]
\begin{center}
\includegraphics[width=1\linewidth]{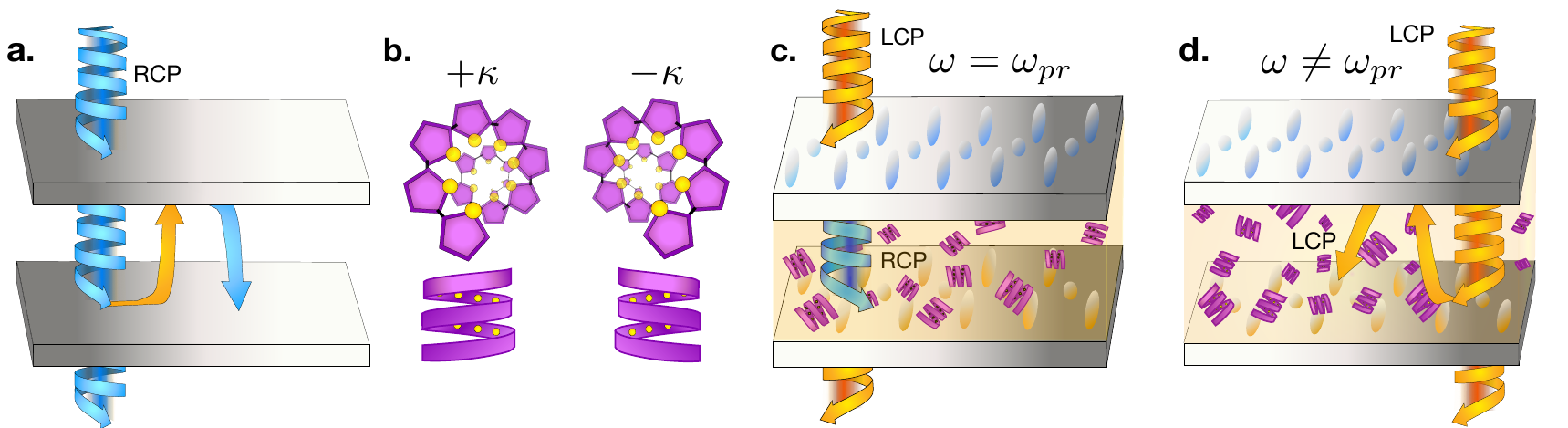}
\caption{%
a) Sketch of a standard optical Fabry-P\'erot cavity made of two silver mirrors.
The polarization is reversed at each metallic mirror reflection inside the cavity.
b) Sketch of chiral thiophene polymers embedded inside the cavity, exhibiting a 3D helical spatial structure~\cite{chiralthiophene}. The molecular optical response in the visible can be tuned by chemical functionalization of the helix side chains.
c) Helicity-preserving Fabry-P\'erot cavity, made of two helicity-preserving dielectric photonic crystal mirrors inspired by Refs.~\cite{Maruf2020,voronin2021single}, filled with the chiral molecules in b).
%
%
%
The dominant process at $\omega=\omega_{pr}$ is the conversion of polarization upon transmission through each PC mirror interface. As such, the helicity-preserving mode cannot be directly driven from the outside of the cavity by LCP\@.
d) Same system as in c) at $\omega$ close but not equal to $\omega_{pr}$. 
The main process is the helicity-preserving reflection inside the cavity. 
Here, the HP mode is coupled to the outside of the cavity and can be driven by LCP light.
%
%
}
\label{Fig1}
\end{center}
\end{figure*}
%
%
%

%
%
An alternative way to enhance OA and CD signals is to increase the optical path and thus the resulting differential dephasing and damping accumulated between left circularly polarized (LCP,$+$) and right circularly polarized (RCP,$-$) waves propagating inside the chiral medium.
At a first glance, the use of a Fabry-P\'erot (FP) cavity~\cite{Born1980,Silverman1994} for this purpose is appealing, due to the possibility of fine control of the cavity optical properties.
Yet, the polarization of a circularly polarized wave is reversed upon each normal reflection at a mirror interface~\cite{jackson1999classical}, thus resulting in a suppression of any CD enhancement~\cite{Corbaton2020,Baranov2020}.
%
%
%
%
To cure this problem, recent works on FP cavities adopted the strategies to use birefringent mirrors~\cite{piao2015spectral}, Faraday rotators~\cite{Sun2022}, metasurfaces~\cite{Yoo2015,Corbaton2020-2,Cao2020}, photonic crystal mirrors~\cite{Maruf2020,voronin2021single}, or to place a planar chiral seed inside the cavity~\cite{Gautier2021}.
%
%
%
%
Some of these works reported up to one~\cite{Sun2022} or two~\cite{Corbaton2020-2} orders of magnitude cavity-induced enhancement of the measured chiroptical signals.
%
%
%
%
Chiral FP cavities are expected to have interesting properties~\cite{hubener2021engineering}, especially upon reaching the polaritonic electronic strong-coupling regime, already observed in normal FP cavities~\cite{ebbesen2016hybrid, schwartz2011reversible,delpo2020polariton}.
For instance, the spontaneous emission rates of LCP and RCP photons by single chiral dipoles inside the cavity were predicted to be different~\cite{voronin2021single}, and the formation of a chiral polariton was reported for cavities with Faraday rotators~\cite{Sun2022}.
%
%

%
%
However, the questions of understanding the physical mechanisms responsible for CD enhancement, and of how to distinguish and optimize the contributions of the mirror, embedded material and chiral polariton to the measured CD, are still open problems.
%

%
%
In this Letter, we propose a theoretical model of a chiral FP cavity made of two helicity-preserving (HP) dielectric photonic-crystal (PC) mirrors. 
%
%
%
%
%
%
%
%
%
For an LCP (RCP) wave entering the cavity, we compute and analyze the total transmission factor $T_{+}$ ($T_{-}$) of the light signal going out of the cavity.
The CD signal is quantified by the differential circular transmission (DCT) as $\mbox{DCT}\equiv 2(T_{+}-T_{-})/(T_{+}+T_{-})$.
By fine tuning of the cavity properties, we show the possibility to enhance the intrinsic contribution to DCT of a chiral Pasteur medium embedded inside the cavity by two orders of magnitude, thus making such an optical device a suitable candidate to realize chiral-discrimination measurements.
We also report the first clear signature of \textit{chiral polariton} formation upon entering the light-matter strong-coupling regime, in the absence of any applied static magnetic field.
%

%
%
%
%
%
\textit{Modelling a FP cavity.---} We first model a normal FP cavity  made of two highly reflective silver mirrors (see \autoref{Fig1}-a)).
Such mirrors are described by a thin homogeneous layer, with Drude-Lorentz permittivity accounting for the metal dispersion and losses~\cite{jackson1999classical,Johnson1972}.
%
%
%
We then fill the cavity with a thiophene polymer medium (see \autoref{Fig1}-b)), that can be functionalized such that the polymer 3D structure is arranged in helices, and has a bright absorption peak in the visible at frequency $\omega_0 \approx 1.9\mbox{ eV}/\hbar $~\cite{chiralthiophene}.
Optical properties of the resulting HLI Pasteur medium, are described by the Condon constitutive relations~\cite{Condon1937}, linking the displacement vector $\vec{D}$ and magnetic induction $\vec{B}$ to the electric field $\vec{E}$ and magnetic field $\vec{H}$:
\begin{eqnarray}
\vec{D} &=& \varepsilon_0 \varepsilon \vec{E} + i\frac{\kappa}{c} \vec{H}
\, ,\label{ConstRel1} \\
\vec{B} &=& \mu_0 \mu \vec{H} - i\frac{\kappa}{c} \vec{E}
\, ,\label{ConstRel2}
\end{eqnarray}
where $c$ is the speed of light and $\varepsilon_0$ ($\mu_0$) is the vacuum permittivity (permeability).
The relative permeability $\mu$ is taken equal to $1$, since the medium is non-magnetic.
The relative permittivity $\varepsilon$ is described by a broad Lorentzian absorption lineshape~\cite{jackson1999classical}:
\begin{equation}
\varepsilon\left(\omega\right) = \varepsilon_{\infty}+\frac{\omega_{p}^2f}{\omega^{2}_{0}-\omega^{2}-i\gamma\omega}
\, ,\label{ConstRel3}
\end{equation}
where $\hbar\gamma=0.1 \mbox{ eV}/\hbar$ is the material damping factor, and $\varepsilon_{\infty}$ is the background relative permittivity.
In \autoref{ConstRel3}, $\omega_{p}$ is the medium plasma frequency (proportional to the square root of the molecular concentration), and $f\in\lbrack 0,1\rbrack$ the oscillator strength for the electronic transition.
The Pasteur coefficient $\kappa$ is fixed at a constant typical value $\kappa = +(-) 10^{-3}$ for left (right)-handed polymer helices.
%
%
Knowing the optical properties of each layer composing the FP cavity, we compute its scattering S-matrix~\cite{carminati2000reciprocity,drezet_reciprocity_2017}, connecting the  input $\vec{E}$ and $\vec{B}$ fields to the output ones (see Supplemental Material~\cite{SupMat}).
%
%
Our system is reciprocal (being linear and without static magnetic field), and its S-matrix is thus symmetric~\cite{carminati2000reciprocity,drezet_reciprocity_2017},
but not unitary, due to the losses in the Pasteur medium and metallic mirrors.
We implemented a transfer-matrix numerical code~\cite{Thesis,SupMat} adapted for layered chiral optical media~\cite{Jaggard1992}, to compute the transmission factors $T_+$ and $T_-$.
%
%
%
\begin{figure*}[tb]
\begin{center}
\includegraphics[width=\linewidth]{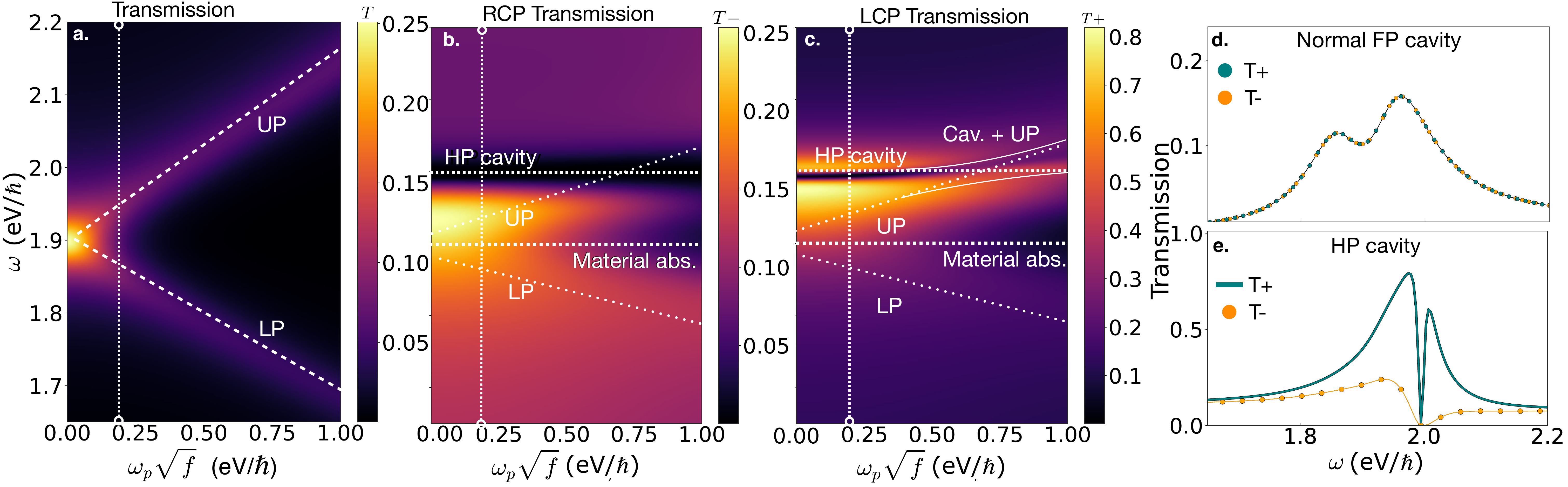}
\caption{%
a) Total transmission factor $T=\left( T_+ + T_-\right)/2$ at normal incidence as a function of incoming light frequency $\omega$ and light-matter coupling strength $\omega_{p}\sqrt{f}$, for a normal FP cavity filled with a 250 nm layer of chiral thiophene polymer resonant at $\omega_0= 1.9\mbox{ eV}/\hbar$.
The polymer linewidth is taken as $\gamma=0.1\mbox{ eV}/\hbar$.
The white dotted lines serve as guidelines for the eye.
b) Transmission map of $T_-\left(\omega,\omega_{p}\sqrt{f}\right)$ for an incoming RCP wave propagating across an helicity-preserving FP cavity, centered at $\omega_{pr} = 2.0 \mbox{ eV}/\hbar$, and with cavity linewidth taken as $0.05 \mbox{ eV}/\hbar$.
The cavity is filled with a 150 nm layer of chiral thiophene.
c) Transmission map of $T_+\left(\omega,\omega_{p}\sqrt{f}\right)$ for an incoming LCP wave, for the same cavity as in b).
d) Plot of $T_-$ (orange dotted) and $T_+$ (green dotted) curves as a function of $\omega$, cut at $\omega_{p}\sqrt{f}=0.2\mbox{ eV}/\hbar$ along the dashed vertical lines in a).
e) Same plot as in d) of $T_-$ (orange dotted line) and $T_+$ (green full line) but for the cuts along the dashed vertical lines in b) and c).
The list of parameters is found in Supplemental Material~\cite{SupMat}.
}
\label{Fig2}
\end{center}
\end{figure*}
%
%

%
%
\textit{Helicity-preserving mirrors.---}%
%
%
%
%
%
%
We complement our approach with a minimal model of PC mirrors, made of two silicon nitride  dielectric layers after Ref. \cite{Maruf2020} (see \autoref{Fig1}-c)).
%
%
The losses in the dielectric mirrors are negligible in the considered frequency range, thus implying for the latter a time-reversal-symmetric (unitary)~\cite{carminati2000reciprocity,Carminati:98,drezet_reciprocity_2017} S-matrix. 
%
%
For the upper (blue) mirror in \autoref{Fig1}-c), the forward transmission matrix $\hat{T}_{\rightarrow}$ and backward reflection matrix $\hat{R}_{\leftarrow}$ are written in the basis of LCP and RCP polarization~\cite{Jones1941} (See Supplemental Material~\cite{SupMat} for a detailed derivation):
\begin{equation}
\hat{T}_{\rightarrow}=
\left(
\begin{array}{cc}
i|t| & 0\\
|t_{+-}|e^{i\psi} & i|t|\\
\end{array}\right);
\mbox{   }
\hat{R}_{\leftarrow}=\left(
\begin{array}{cc}
0 & -i|t|e^{i2\psi}\\
-i|t|e^{i2\psi} & r_{--}\\
\end{array}\right),
\label{jonesmatrices}
\end{equation}
where $|t|=\sqrt{\frac{1-|t_{+-}|^{2}}{2}}$, $r_{--}=-|t_{+-}|e^{i\psi}$ and $|t_{+-}|e^{i\psi}=\gamma_{pr}/\left[{i\left(\omega-\omega_{pr}\right)+\gamma_{pr}}\right]$ is the complex cross-transmission factor~\cite{Gorkunov2020,Kondraton2016,Fan2003}.
%
%
In the band centered at the frequency $\omega_{pr}=2.0 \mbox{ eV}/\hbar$ with bandwidth $\gamma_{pr}= 50 \mbox{ meV}/\hbar$, \autoref{jonesmatrices} enables to reproduce the frequency-dependence of all transmission and reflection coefficients of the HP mirror reported in Ref.~\cite{Maruf2020} and modelled in Ref.~\cite{voronin2021single}. 
For $\omega\approx\omega_{pr}$, this mirror acts as an efficient polarization cross-converter in transmission for incoming LCP waves ($|t_{-+}|=0,|t_{+-}|\approx 1$) and reflector for RCP waves ($|r_{++}|=0,|r_{--}|\approx 1$).

%
%
Analogously to the definition of helicity-content in the context of duplex-symmetry for free-space \cite{Barnett_2012}, we investigate if the mirror involves helicity-preservation in transmission or reflection. 
%
%
%
To this aim, we examine the ratio between $|t_{+-}|$ and $|t|$. 
When $|t|\ll|t_{+-}|$, we find the case of an ideal HP mirror, for which the helicity of RCP light is  conserved in reflection and the one of LCP light is cross-converted in transmission. 
%
%
The situation is reversed for $|t|\gg|t_{+-}|$, recovering the case of a normal mirror. 
%
%
%
%
%
%
%
We note that the PC mirror properties for a wave approaching the upper (blue) surface from below are reverted, namely LCP light is reflected while RCP light is cross-converted in transmission.
We are then able to complete the cavity with the lower (orange) mirror in \autoref{Fig1}-c), obtained by flipping the surface of the previous one.
Contrary to the case of normal mirrors, the obtained FP cavity is helicity-preserving~\cite{Corbaton2016,voronin2021single}.
Indeed, internal reflections are allowed only for incoming LCP waves, whose polarization is preserved upon reflection (see \autoref{Fig1}-d).
However, the same polarization (LCP) that is efficiently reflected internally is also the one which gets cross-converted upon transmission, namely from LCP to RCP (see \autoref{Fig1}-c).
As a consequence, when the mirrors act as perfect cross-converters and reflectors $\omega = \omega_{pr}$ (\autoref{Fig1}-c), the circularly polarized cavity mode is decoupled from the outside of the cavity, and cannot be populated efficiently from the outside with incoming LCP waves (see Supplemental Material~\cite{SupMat}).
On the other hand, when $\omega$ is detuned with respect to $\omega_{pr}$ (\autoref{Fig1}-d), the HP mode couples efficiently to the outside of the cavity. 
%
%
%
We thus choose a material absorbing at $\omega_0 = 1.9 \mbox{ eV}/\hbar\neq \omega_{pr}$, and drive it at frequencies detuned with respect to $\omega_{pr}$.
%
%
%
%
In this way, we can observe the effect of the HP cavity in a spectral region which does not overlap with the transparency window of the cavity.
%

%
%

%
\textit{Cavity optical transmission.---}%
We show in \autoref{Fig2}-a), the 2D map of the average transmission $T=\left( T_+ + T_- \right)/2$ across a normal FP cavity, as a function of the input light frequency $\omega$ and light-matter coupling strength $\omega_{p}\sqrt{f}$.
Light enters the cavity at normal incidence, and one of the cavity optical modes is brought into resonance with the material absorption frequency $\omega_0$.
The plots of $T_{+}(\omega)$ and $T_{-}(\omega)$ corresponding to the cut along the dashed white line at $\omega_{p}\sqrt{f}=0.2 \mbox{ eV}/\hbar$, are presented in \autoref{Fig2}-d) as blue and orange dots respectively.
We obtain as expected that the $T_{+}$ and $T_{-}$ signals are not significantly different (the difference is actually two orders of magnitudes smaller than $\kappa$), which confirms that the normal cavity does not enhance the CD signal.
This remains true even upon entering the polaritonic strong-coupling regime~\cite{Zhu1990}, for which lower (LP) and upper (UP) polaritonic branches appear (see \autoref{Fig2}-a)).
This is due to the above mentioned polarization-reversal of the waves upon each reflection at the metallic mirrors.
%
%
%
%
%
%
%

%
The 2D maps for $T_{+}$ and $T_{-}$ obtained in the case of an HP cavity (see \autoref{Fig1}-d)), are shown in \autoref{Fig2}-b) and \autoref{Fig2}-c) respectively.
They are strikingly different from the case of normal FP cavity. 
As shown in \autoref{Fig2}-e), there exists a band at $\omega=\omega_{pr}$ for which the LCP wave is perfectly transmitted whereas there is no transmitted RCP wave, as it is completely reflected by the upper mirror.
%
%
The transmission of an LCP wave is the superposition of two distinct effects: the first is the cross-conversion of polarization between LCP$\rightarrow$RCP$\rightarrow$LCP (with efficiency $1$ at $\omega = \omega_{pr}$) as light propagates across the structure.
This corresponds to a full transparency of the cavity to the LCP driving. 
Since the incoming LCP field is fully converted to RCP, the standing-wave LCP mode inside the cavity is fully decoupled from the outside of the cavity, similar to a Bound State in the Continuum~\cite{BIC}.
%
%

%
The second effect is observable on the red-shifted tail of the cavity resonance.
%
%
Here, the behavior of the cavity deviates from the ideal cross-conversion and perfect reflection, allowing leakage of LCP inside the cavity mode.
As such, the UP of the material couples efficiently to the circularly polarized cavity mode, which exhibits a Fano-like lineshape. The Fano-like feature stems from the interference between the narrow (localised) helicity-preserving mode and the spectrally broad achiral Fabry-P\'erot modes and material absorption that act as an effective continuum \cite{piao2015spectral}.
As a consequence, the HP cavity significantly modifies the CD signal in the proximity of the HP band.
Consequently, the UP acquires an imbalance between its left and right polarization content upon reaching the HP region.
This is the first clear demonstration of a \textit{chiral polariton} excitation in a non-magnetic FP cavity, and one of the main result of this paper.
%

%
%
\textit{DCT signals.---}%
%
%
\begin{figure}[tb]
\begin{center}
\includegraphics[width=1\linewidth]{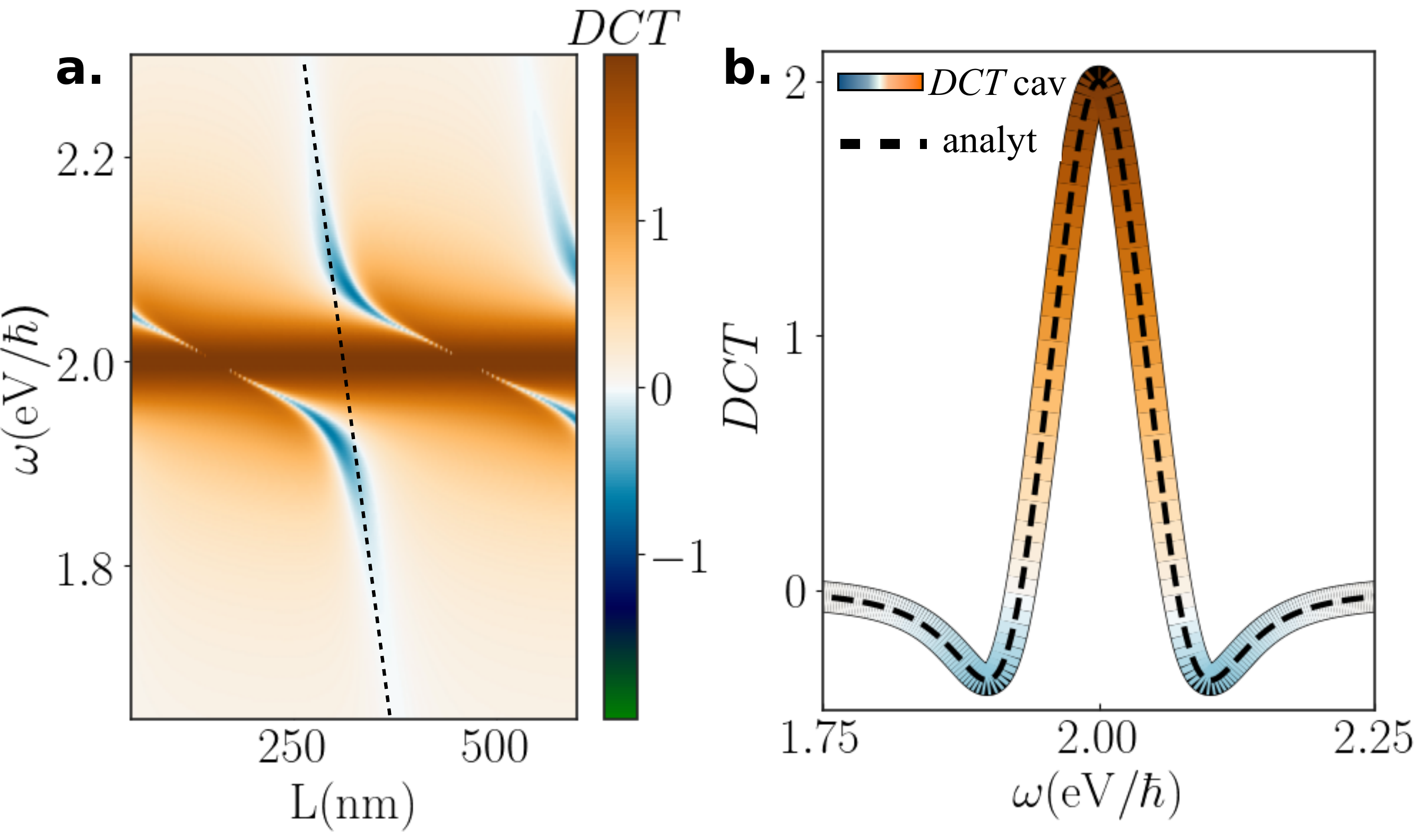}
\caption{%
a) 2D map of DCT signal as a function of incoming light frequency $\omega$ and cavity length $L$, in an HP cavity.
The optical cavity modes are shown as blue areas in the proximity of the HP band appearing in orange.
b) Plot of $DCT(\omega)$ realized along the dashed black line in panel a) corresponding to the cavity light mode for $L=\lambda/2$.
The comparison with the analytical result of \autoref{DCTchiralMir} is shown as a dashed black line.
}
\label{Fig3}
\end{center}
\end{figure}
%
In order to interpret the previous results, and unravel the role of dielectric PC mirrors in the mechanism of CD enhancement, we show in \autoref{Fig3}-a), the 2D map for DCT as a function of incoming light frequency $\omega$ and cavity length $L$.
%
%
This signal is computed for an empty HP cavity, namely in absence of embedded Pasteur medium.
The HP band appears as a brown non-dispersive region, whereas the dispersive optical cavity mode is seen as a blue region.
%
%
%
Narrow asymmetric peaks~\cite{Fan2003,piao2015spectral} appear, which are generated by interaction between the broad cavity-mode tail and the narrow HP mode.

We present and analyze in more detail in \autoref{Fig3}-b) the DCT signal for a cut along the cavity optical mode shown as a dashed black line in panel a).
We find an exact analytical expression for this signal~\cite{SupMat}:
\begin{equation}
DCT = \frac{16(4-\delta^{4})}{32\left(1+\delta^{2}\right)+8\delta^{4}+\delta^{8}} \, ,
\label{DCTchiralMir}
\end{equation}
%
%
with $\delta \equiv \left(\omega-\omega_{pr}\right)/\gamma_{pr}$ the relative detuning with respect to the HP band.
%
%
%
The outcome of \autoref{DCTchiralMir} is shown as a dashed black line in \autoref{Fig3}-b), and perfectly matches our transfer-matrix numerical calculation.
At the center of the HP band $(\delta = 0)$, the PC mirrors are designed to let the LCP (RCP) wave be perfectly transmitted (reflected) by the cavity, with $T_+=1$ ($T_-=0$), so that $DCT=2$, thus realizing a point of duality symmetry and maximum electromagnetic chirality~\cite{Corbaton2016}.
%
%
%
%
%
%
%
%
%
Yet, as commented previously, the cavity mode at $\delta = 0$ is decoupled from the outside of the cavity, limiting its utility for chiral sensing as the field and thus the light-matter interaction inside the cavity are not enhanced.
%
%
More interesting features appear away from the HP region ($|\delta| \gg 1$), where the DCT gets negative and decays with the fourth power of inverse detuning $DCT \approx - 16/\delta^4$.
%
%
The vanishing of DCT at large detuning simply recovers the behavior of normal mirrors.
The remaining negative DCT in this region arises from the imbalance between the reflection factors $r_{--}$ and $r_{++}$ at the mirror interface (see \autoref{jonesmatrices}), which tends to increase $T_-$ with respect to $T_+$.
%
%
%
%
\begin{figure*}[tb]
\begin{center}
\includegraphics[width=1\linewidth]{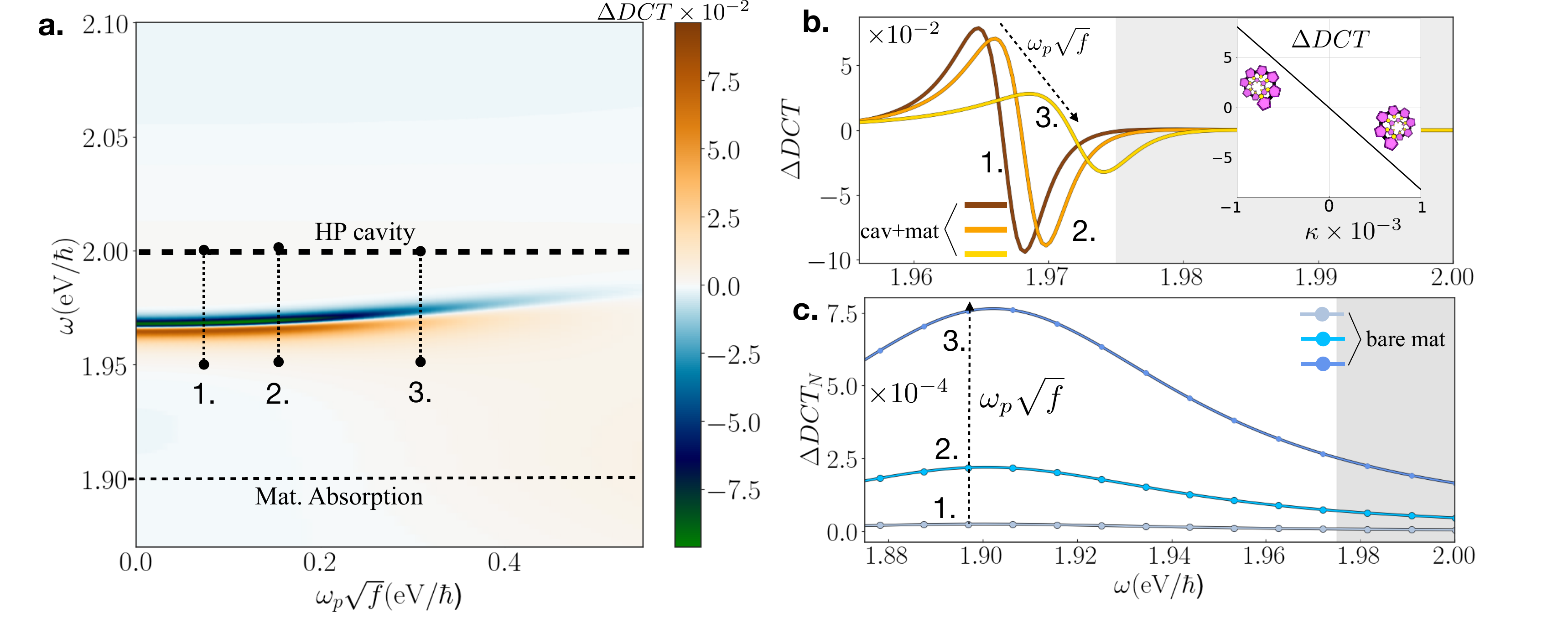}
\caption{
a) Map of $\Delta DCT$, as a function of input light frequency $\omega$ and light-matter coupling strength $\omega_{p}\sqrt{f}$, for the HP cavity of \autoref{Fig2}-b) and \autoref{Fig2}-c), filled with a $150 \mbox{ nm}$ thick layer of the same Pasteur medium absorbing at 1.9 eV/$\hbar$.
%
%
b) Cuts of $\Delta DCT(\omega)$ for given $\omega_{p}\sqrt{f}$ ($0.05$ brown, $0.15$ orange and $0.30$ yellow respectively), along the dashed black lines of pannel a).
The grey area represents the HP band.
The variation of $\Delta DCT$ at $\omega=1.967 \mbox{ eV}/\hbar$ is shown as a function $\kappa$ in the inset of panel b).
c) Same plots of the DCT-enhancement labelled $\Delta DCT_N(\omega)$, but in the case of the normal FP cavity of \autoref{Fig2}-a).
The increase in the oscillator strength corresponds to a more absorptive material at fixed chiral material thickness (no role of the cavity normal mirrors).
%
%
}

\label{Fig4}
\end{center}
\end{figure*}
%
%
%
Upon coming closer to the HP region, the DCT signal decreases, finally reaching a minimum and large negative value close to the crossover (located at $|\delta| = \sqrt{2}$) between the normal and HP regions, where it becomes possible to populate directly the helicity-preserving mode.
The existence of such a region of strong negative DCT, is another important result of this paper: it is associated with an asymmetry in polarization content of the electromagnetic field stored inside the cavity (see Supplemental Material~\cite{SupMat}).
The presence of such regions are promising to achieve enhanced chiral sensing,
and is a general feature of the modeled cavity, as shown by \autoref{DCTchiralMir}.
In addition, these regions allow to reveal the emergence of the chiral polaritons reported in \autoref{Fig2}: by increasing $\omega_{p}\sqrt{f}$, the position of the UP enters the blue region, and as a result the chiral material couples efficiently with the circularly polarized cavity mode.
%

%
%
\textit{Chiral sensing.---}%
To reveal the consequences of this coupling with the cavity mode, we reconsider the HP cavity filled with the Pasteur medium of \autoref{Fig1}-d), and investigate the $\Delta DCT$~\cite{Thesis}, measured as:
\begin{equation}
\Delta DCT(\kappa)=DCT(\kappa)-DCT(0)
\label{DeltaDCT} \, .
\end{equation}
This signal enables to subtract the dominant background contribution of the HP mirrors (shown in \autoref{Fig3}) obtained at $\kappa=0$ from the $\kappa$-dependent DCT\@.
We present in \autoref{Fig4}-a) the 2D map of $\Delta DCT$ corresponding to the same cavity as in \autoref{Fig2}-b) and c), except for the thickness of the Pasteur medium chosen at $150 \mbox{ nm}$.
As expected, the HP band and polariton branches are absent from this map, since they are part of the removed background signal.
%
%
Remarkably, the crossover region experiencing an enhanced negative DCT is still observed in \autoref{Fig4}-a), as a blue region with $\Delta DCT<0$.
In this region a thin brown asymmetric resonance emerges (also seen in \autoref{Fig2}-c), which presents a strong positive $\Delta DCT$.
We further characterize in \autoref{Fig4}-b), the frequency dependence of this resonance cut along three dashed black lines in \autoref{Fig4}-a), each corresponding to different values of $\omega_{p}\sqrt{f}$.
We show that in the orange region where the material couples to the cavity mode, the corresponding $\Delta DCT$ signal is up to two orders of magnitude higher than the corresponding signals obtained in the case of a normal FP cavity (blue curves labelled as $\Delta DCT_N$ in \autoref{Fig4}-c).
Moreover, as shown in the inset of \autoref{Fig4}-b), the DCT reverses sign upon changing the medium chirality (and thus the sign of $\kappa$).
This is a clear proof that the $\Delta DCT$-signal is a measure of the intrinsic chirality of the molecules embedded inside the cavity.
%
%

%
As a last important result of this paper, we predict the possibility to perform chiral sensing experiments by fine optical tuning of a non-magnetic HP cavity, \textit{with two orders of magnitude enhancement} compared to a normal cavity.
The inset of \autoref{Fig4}-b) shows indeed a linear dependence with $\kappa$ of $\Delta DCT(\kappa) \approx \mathcal{S} \kappa$, with $\mathcal{S}$ being the slope or sensitivity of this measurement.
This suggests to use this linear signal for calibrating the excess concentration in left or right chiral molecules in a given mixture of enantiomers.
Finally, we see in \autoref{Fig4}-b) that for the HP cavity, $\Delta DCT$ decreases with the polaritonic coupling $\omega_{p}\sqrt{f}$, while the opposite is true for the normal cavity of \autoref{Fig4}-c).
This is due to a modification in the impedance mismatch between the Pasteur layer and the mirrors, thus resulting in a $\omega_{p}\sqrt{f}$-dependent shift of the blue region~\cite{SupMat}.
%

%
%
%
In conclusion, we have investigated chiroptical properties of a non-magnetic helicity-preserving Fabry-P\'erot cavity, made of two dielectric photonic crystal mirrors.
We have shown the existence of a region close to the helicity-preserving band where the material can couple efficiently to the circularly polarized cavity mode.
As a result of this mechanism, we predict that the intrinsic contribution to circular dichroism of a chiral Pasteur medium can be enhanced by up to two orders of magnitude compared to a normal Fabry-P\'erot cavity.
We show the promising possibility for biological, chemical and pharmaceutical applications, to use this cavity for chiral-sensing measurements.
Finally, we unravel the presence of chiral polariton excitation, that can be generated upon reaching the electronic strong-coupling regime.
We hope that this work will stimulate new experiments and open new roads to investigate chemical reactivity of materials in cavity-confined chiral optical environments~\cite{FJ2018,Li2021,Sun2022}.
%
%

%
%
\section*{Acknowledgments}
\label{Acknowledgments}
L. Mauro and R. Avriller acknowledge financial support by Agence Nationale de la Recherche project CERCa, ANR-18-CE30-0006, EUR Light S\&T Graduate Program (PIA3 Program “Investment for the Future”, ANR-17-EURE-0027), IdEx of the University of Bordeaux / Grand Research Program GPR LIGHT, and Quantum Matter Bordeaux.
J. Fregoni and J. Feist acknowledge financial support by European Research Council through Grant ERC-2016-StG-714870 and by Spanish Ministry for Science, Innovation, and Universities – Agencia Agencia Estatal de Investigación through Grants No. RTI2018-099737-B-I00 and PID2021125894NB-I00,
and CEX2018-000805-M (through the María de Maeztu Program for Units of Excellence in R\&D).
%
%
\bibliography{biblio}

\end{document}